\documentclass[lettersize,journal]{IEEEtran}
\usepackage{amsmath,amsfonts}
\usepackage{algorithmic}
\usepackage{algorithm}
\usepackage{array}
\usepackage{xcolor}
\usepackage[T2A,T1]{fontenc}
\usepackage{textcomp}
\usepackage{stfloats}
\usepackage{url}
\usepackage{verbatim}
\usepackage{graphicx}
\usepackage{textcomp}
\usepackage{xcolor}
\usepackage{cite}
\usepackage{hyperref}
\usepackage{lettrine}
\usepackage{comment}
\usepackage{caption}
\usepackage{amsmath,amssymb,amsfonts}
\usepackage{subcaption}
\usepackage{bm}
\def\BibTeX{{\rm B\kern-.05em{\sc i\kern-.025em b}\kern-.08em
    T\kern-.1667em\lower.7ex\hbox{E}\kern-.125emX}}
\usepackage{balance}
\begin{document}
\title{Hybrid FIM and STAR-BD-RIS-Aided Wireless Communications with Short Packet Length: A Meta-TD3 Approach}
\author{Ayla Eftekhari, Maryam Cheraghy, Armin Farhadi, Mohammad Robat Mili, and Qingqing Wu,~\IEEEmembership{Senior Member, IEEE}, \thanks{ A. Eftekhari is with the Biomedical Engineering of Amirkabir University of Technology, Tehran, Iran (e-mail: a-eftekhari.kh@aut.ac.ir). M. Cheraghy is with the Department of Computer Science, Wenzhou-Kean University, Wenzhou 325060, China, and also with the Department of Computer Science and Technology, Kean University, Union, NJ 07083 USA (e-mail: mcheragh@kean.edu). A. Farhadi is with the School of Electrical and Computer Engineering, College of Engineering, University of Tehran, Tehran, Iran (e-mail: armin.farhadi@ut.ac.ir). M.R. Mili is with Pasargad Institute for Advanced Innovative Solutions (PIAIS), Tehran, Iran, (e-mail: amirhosein.mohammadzadeh@piais.ir, mohammad.robatmili@piais.ir). Q. Wu is with Shanghai Jiao Tong University, 200240, China (e-mail:
					qingqingwu@sjtu.edu.cn).}}

\markboth{}%
{How to Use the IEEEtran \LaTeX \ Templates}

\maketitle

\begin{abstract}
Reconfigurable intelligent surfaces (RIS) and flexible intelligent metasurfaces (FIM) have been widely adopted in multi-user wireless communication systems to enhance channel quality through simultaneous transmission and reflection of signals and three-dimensional reconfiguration of antennas. In this paper, we propose a novel system architecture that integrates the benefits of both technologies by deploying an FIM antenna at the base station (BS) and a simultaneously transmitting and reflecting beyond diagonal RIS (STAR-BD-RIS) along the transmission path to ensure sufficient received power for single-antenna users. The objective is to maximize the achievable sum rate considering the short block length by jointly optimizing the FIM surface configuration, the transmit beamforming vector, and STAR-BD-RIS phase shift matrix subject to practical constraints including minimum signal-to-interference-plus-noise ratio (SINR), power limitations, FIM constraint, and the STAR-BD-RIS phase-shift matrix. To solve the resulting non-convex optimization problem, we develop a learning-based approach that incorporates meta-learning into the twin delayed deep deterministic policy gradient (TD3) algorithm, referred to as Meta-TD3. The simulation results demonstrate that the proposed hybrid system outperforms conventional configurations employing either FIM or RIS alone, while the Meta-TD3 algorithm achieves superior performance compared to classic learning techniques.
\end{abstract}
\vspace{-1pt}

\vspace{-16pt}
\section{Introduction}
\vspace{-5pt}
The 5G and upcoming 6G systems aim to support advanced applications such as internet of things (IoT), brain–computer interfaces, robotics, and health monitoring \cite{farhadi2024meta,farhadi2025joint,de2025gwo}. These demand compact antennas and high quality of service (QoS) to ensure terabit-level data rates, ultra-reliable connectivity, and microsecond latency for large-scale deployments \cite{javadi2025meta}. To achieve this, advanced antenna technologies in terahertz and millimeter-wave bands are essential for ultra-reliable low-latency communication (URLLC) and enhanced broadband, while reducing energy use and hardware cost \cite{Ziolkowski2022,farhadi2025joint}.

Reconfigurable intelligent surfaces (RIS) and flexible intelligent metasurfaces (FIM) provide energy-efficient, spectrum-optimized solutions for next-generation wireless systems. Unlike fixed antennas, they dynamically adapt to channel variations and reconfigure multi-input multi-output array geometry within a defined region \cite{wu2024movable}.
RISs consist of passive elements that independently control the phase and amplitude of incident electromagnetic waves. Their extensions include the beyond-diagonal RIS (BD-RIS), which enables full-space signal control to enhance spatial coverage and beam directivity, especially in multi-user systems \cite{10892200}, and the simultaneously transmitting and reflecting RIS (STAR-RIS), which supports bidirectional operation to extend coverage beyond the reflection area \cite{10892200}. Building on these, the STAR-BD-RIS architecture integrates the strengths of both, offering fine-grained control of wave propagation in transmission and reflection domains, making it highly suitable for dynamic wireless environments \cite{10892200}. In parallel, FIMs introduce physical reconfigurability into transceiver design. Typically fabricated on soft substrates such as polydimethylsiloxane \cite{Kamali}, they consist of dense radiating element arrays whose positions can be tuned through surface morphing. This enables adaptation to desired 3D shapes while preserving electromagnetic properties, allowing a constructive combination of multipath components at the receiver and thus improving channel capacity and communication quality \cite{An2025}.


However, previous works have mainly studied STAR-BD-RIS \cite{10892200}, FIM antennas \cite{An2025}, or alternative designs such as rotating RISs \cite{cheng2022ris}, movable RISs with passive beamforming \cite{zhang2024ris}, and FIMs combining passive beamforming with reconfiguration \cite{yang2025flexible} independently. In addition, the potential gain of deploying these systems with short block transmissions to maximize the achievable rate and supporting reliable communication for multiple users has not yet been explored in the literature. In contrast, this paper proposes a hybrid architecture that integrates FIM at the base station (BS) with STAR-BD-RIS in the environment. By combining these two technologies and employing short block length transmissions to meet stringent latency requirements, the proposed approach improves beamforming precision, increases spatial degrees of freedom, and enhances achievable rate, making it a promising solution for 6G networks. Hence, the proposed optimization problem is highly non-convex and involves many coupled parameters. In such scenarios, conventional optimization methods often fail to provide reliable real-time solutions.
To address the resulting complex optimization, we propose a meta-twin delayed deep deterministic policy gradient (Meta-TD3) framework that jointly optimizes FIM configuration, transmit beamforming, STAR-BD-RIS phase shifts, and per user signal-to-interference-plus-noise ratio (SINR) constraints to maximize achievable rate. TD3 is chosen for its reduced overestimation bias and stability, while integration with meta-learning improves generalization, accelerates convergence, and enhances long-term performance.
The main contributions of this paper can be summarized as follows:
\begin{itemize}
    \item We introduce a novel system architecture that combines FIM at the BS with STAR-BD-RIS, enabling conflict with channel situations and enhanced coverage.
    \item We formulate a joint optimization problem considering the short block length to optimize to maximize the achievable rate for each user. In the defined optimization problem, we are going to optimize the problem by considering the minimum SINR for each user, the maximum available power at the BS, FIM, and STAR-BD-RIS constraints.
    \item The investigated optimization problem is a type of NP-hard optimization problem, which cannot be solved with conventional methods. Hence, we investigate a Meta-TD3 algorithm to efficiently solve the highly non-convex problem by considering a proper strategy to train the meta-deep reinforcement learning (meta-DRL) algorithm. 
\end{itemize}
\vspace{-11pt}
 
\section{System and Signal Model}
\vspace{-5pt}
\subsection{System Model}

In this article, we propose a system model for achieving complete signal coverage in space, where a BS equipped with FIM technology communicates with \( N \) single-antenna mobile users. These users are categorized into two groups: those located in the transmission area and those in the reflection area. Accordingly, the set of users, denoted by \( \mathcal{N} \), is defined as \(\mathcal{N} = \{ 1, \dots, N_{s_t}, N_{s_t} + 1, \dots, N_{s_t} + N_{s_r} \},\)
where \( N_{s_t} \) represents the number of users in the transmission sector, and \( N_{s_r} \) represents the number of users in the reflection sector. Each user is denoted by \( n_s \), with \( n_s \in \mathcal{N} \).

\begin{figure}[!t]
\centering
\includegraphics[width=0.55\linewidth, height=5cm]{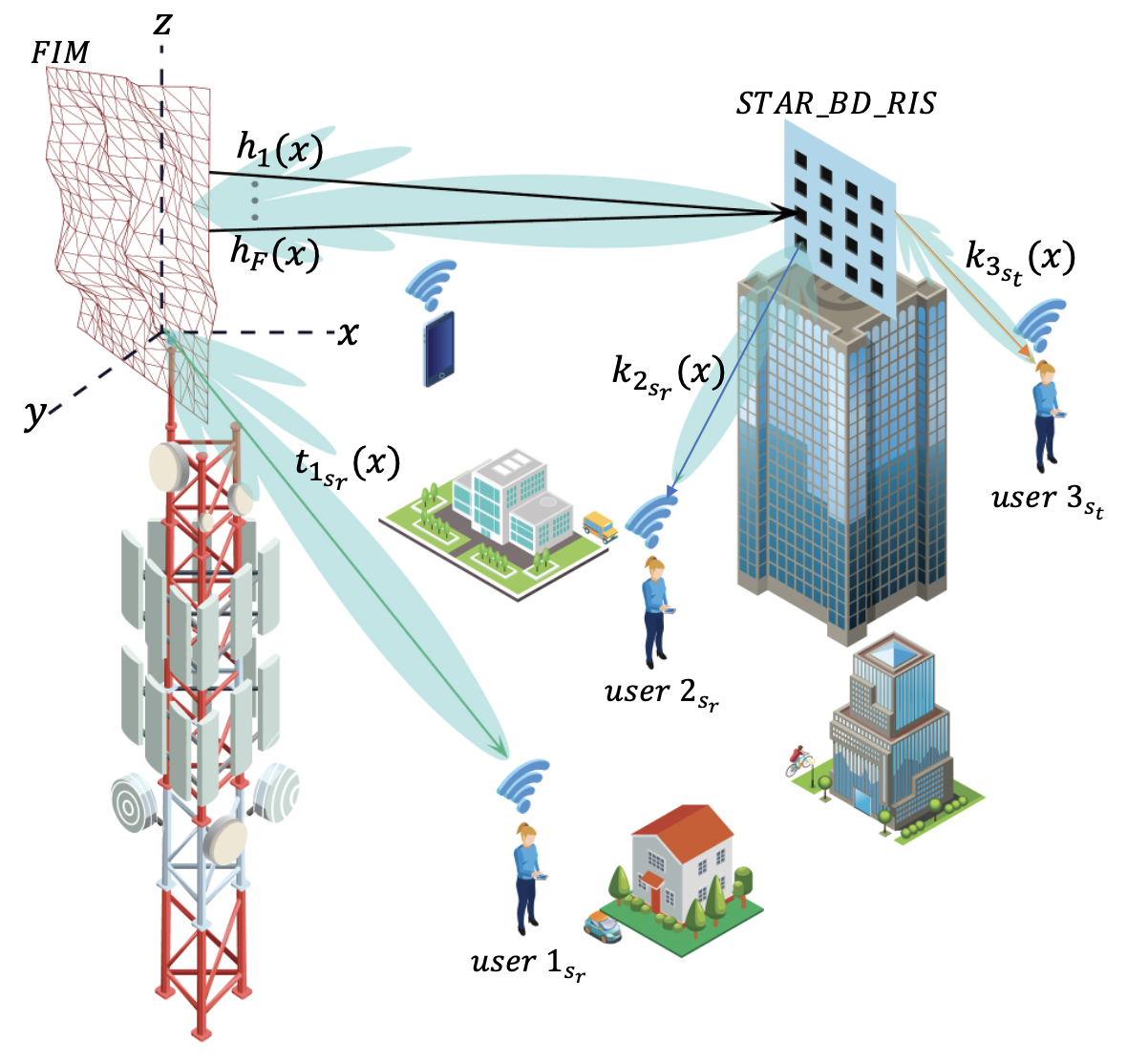}
\caption{\small System architecture of the proposed system with FIM at the BS and STAR-BD-RIS in the transmission path.}
\label{fig1}
\vspace{-15pt}
\end{figure}

However, since the direct link between the BS and users may suffer from poor quality due to various reasons, such as obstacles and buildings, it is assumed that a STAR-BD-RIS is placed along the path between the BS and users. This STAR-BD-RIS assists the FIM antenna in the BS to ensure efficient signal transmission to users. As illustrated in Fig.~\ref{fig1}, we consider the downlink of a multiuser multi-input single-output communication system, where BS utilizes FIM technology to serve multiple users with the help of STAR-BD-RIS, ensuring efficient signal propagation.
\vspace{-10pt}
\subsection{Channel Model}
In BS, the FIM antenna consists of a set of radiating elements indexed by \( \mathcal{P} = \{1, \dots, P\} \), arranged as a uniform planar array (UPA) on the \( y \)–\( z \) plane. The total number of elements is given by \( P = P_y P_z \geq N \), where \( P_y \) and \( P_z \) denote the number of elements along the \( y \) and \( z \) axes, respectively. On the other hand, it is assumed that the STAR-BD-RIS antenna has \( F \) antennas, represented by the set \( \mathcal{F} = \{1, \dots, F\} \).

In the proposed system, each radiating element of the FIM in the BS can be dynamically repositioned along the \( x \)-axis, which is perpendicular to the surface, with the help of a controller. Specifically, the position of the \( p \)-th radiating element is given by
\(
\bm{O}_p = [x_p, y_p, z_p]^T \in \mathbb{R}^3, \quad \forall p \in \mathcal{P}.
\)

These elements are arranged in the \( y \) and \( z \) directions with a predefined number and specific positions. In addition, there is a fixed spacing between consecutive elements. Assuming the first element as the reference point, this spacing is calculated as
\(y_p =r_y\times\text{mod}\,(p-1, P_y),
\) and
\(z_p = r_z \times \lfloor (p-1)/P_z \rfloor, \quad \forall p \in \mathcal{P}, 
\) where \( r_y \) and \( r_z \) represent the spacing between adjacent antenna elements in the \( y \) and \( z \) directions, respectively.
Furthermore, the coordinate \( x_p \) can be flexibly adjusted in the direction perpendicular to the surface through the morphing process within a predefined range, constrained by the reversible deformation capability of the FIM, satisfying the condition
\(0 \leq x_p \leq x_{\text{max}}, \quad \forall p \in \mathcal{P}.
\) where \( x_p \) represents the horizontal position of each radiating element \cite{An2025}.

The transmitted signal from the BS to the user can propagate through multiple paths, including both direct transmission and reflected transmission via STAR-BD-RIS, specifically:

\subsubsection{Direct Transmission}

The signal is transmitted directly from the BS to the user. If we divide the area into two sectors, this partition represents the users in each specific area, defined as the set \( \mathcal{S} = \{s_t, s_r\} \), where \( s_t \) and \( s_r \) denote the transmission and reflection sectors, respectively. The channel vector from the BS to the users in each sector is represented as
\(\bm{t}_{n_s} \in \mathbb{C}^{P \times 1}, \quad \forall n_s \in \mathcal{N}\).
Let \( D \) denote the number of channel paths, and define the corresponding path set as \( \mathcal{D} = \{1, 2, \dots, D\} \). The complex gain of the \( d \) th propagation path from the BS to the user is denoted by
\(
g_{n_s,d} \in \mathbb{C}, \quad \forall d \in \mathcal{D}, \quad \forall n_s \in \mathcal{N}, 
\)
while \( \varphi_d \in [0, \pi) \) and \( \varrho_d \in [0, \pi) \) represent the elevation and azimuth angles of the \( d \) th propagation path, respectively.
Therefore, the narrowband channel \( \bm{t}_{n_s}(\bm{x}) \) can be written as $\bm{t}_{n_s}(\bm{x}) = \sum_{d=1}^{D} g_{n_s,d}~\bm{q}(\bm{x}, \varrho_d, \varphi_d), \quad \forall n_s \in \mathcal{N}, $
where \( \bm{q} (\bm{x}, \varrho, \varphi) \) is expressed as
\begin{align}
\bm{q}(\bm{x}, \varrho, \varphi) &=  
\big[1, \dots ,e^{j \frac{2 \pi}\lambda (x_p \sin \varphi \cos \varrho + y_p \sin \varphi \sin \varrho + z_p \cos \varphi)}, \notag \\  
&\quad \dots, e^{(j \frac{2 \pi}\lambda x_P \sin \varphi \cos \varrho + y_P \sin \varphi \sin \varrho + z_P \cos \varphi)} \big]^T, \notag  
\end{align}
where \( \lambda \) denotes the carrier wavelength.

In this paper, we assume that \( g_{n_s,d} \) independent and identically distributed (i.i.d.) complex Gaussian random variables, with
\(
g_{n_s,d} \sim \mathcal{CN}(0, \xi_{{n_s} , d}^2), \quad \forall n_s \in \mathcal{N},\quad \forall d \in \mathcal{D},
\)
The term \( \xi^2_{n_s,d} \) represents the average power of the signal on the \( d \) th path to user \( n_s \), which is equivalent to the path loss \( L_{n_s} \) between the user \( n_s \) and the BS. This relationship is given by
\(\sum_{d=1}^{D} \xi_{n_s,d}^2 = L_{n_s},\quad \forall n_s \in \mathcal{N}, \)
where \( L_{n_s} \) is the path loss for the \( n_s \)th user.

\subsubsection{Reflected and Transmitted Transmission through STAR-BD-RIS}

The signal is first received by the STAR-BD-RIS, then reflected or transmitted toward the users. In this scenario, the channel vector from the BS to users, has two parts:
\(
\bm{H} \in \mathbb{C}^{P \times F}
\)
denotes the channel matrix connecting the BS to the STAR-BD-RIS, constructed as \(\bm{H}=[\bm{h}_{1}(\bm{x}),\bm{h}_{2}(\bm{x})\dots,\bm{h}_{F}(\bm{x})]\). where each \(\bm{h}_{f}(\bm{x})\) is a superposition of \(D\) propagation path can be expressed as $\bm{h}_{f}\bm{(x}) = \sum_{d=1}^{D} \alpha_{f,d} ~ \bm{q} (\bm{x}, \varrho_{d,f}, \varphi_{d,f}), \quad \forall f \in \mathcal{F}, $
where \( \bm{q} (\bm{x}, \varrho_{d,f}, \varphi_{d,f}) \) is defined as in the previous section, which \(\varrho_{d,f}\), \(\varphi_{d,f}\) represent the elevation and azimuth angles of the \( d \) th propagation path from FIM to the STAR-BD-RIS channel, respectively. \(\alpha_{f,d}\) denotes the complex path gain, which is an independent and identically distributed (i.i.d.) complex Gaussian random variable. 
Similarly to $\bm{t}_{n_s}(\bm{x})$, the channel \(
\bm{k}_{n_s} \in \mathbb{C}^{F \times 1}
\) between the STAR-BD-RIS and user \( n_s \), where the STAR-BD-RIS can be expressed as $\bm{k}_{n_s}(\bm{x}) =\sum_{d=1}^{D} \beta_{n_s,d} ~ \bm{q} (\bm{x}, \varrho_{d,n_s}, \varphi_{d,n_s}), \quad \forall n_s \in \mathcal{N},$
\vspace{-10pt}
\subsection{STAR-BD-RIS Model}
The STAR-BD-RIS is modeled using a two-section scattering matrix. It consists of two independent phase shift matrices, denoted as \( \boldsymbol{\Omega}_s \), which correspond to each section and together define the overall STAR-BD-RIS matrix. These matrices must satisfy the unitary constraint $\sum_{s \in \mathcal{S}}  \boldsymbol{\Omega}_s^H \boldsymbol{\Omega}_s = \bm{I}_F, \quad \forall s \in \mathcal{S},$
where \( \bm{I}_F \) represents the \( F \times F \) identity matrix.

In this study, we utilize a cell-wise single-connected architecture for the STAR-BD-RIS. In this configuration, the phase shift matrices are represented in a diagonal form:
\(
\boldsymbol{\Omega}_s = \operatorname{diag}(\Omega_s^1, \Omega_s^2, \dots, \Omega_s^F), \quad \forall s \in \mathcal{S}, 
\)
where each \( \Omega_s^f \) represents the phase shift of the \( f \) th element. The STAR-BD-RIS constraint can be reformulated as
\(
\left| \Omega_s^f \right|^2 = 1, \quad \forall f \in \mathcal{F}, \quad \forall s \in \mathcal{S}. 
\)
This means that each phase shift element operates independently without interconnection between different elements \cite{farhadi2025meta}.
\vspace{-15pt}
\subsection{Signal Model}

The complex baseband signal received from the user \( n_s \) can be expressed as $S_{n_s} = \left( \bm{k}_{n_s}^H \boldsymbol{\Omega}_s \bm{H}^H + \bm{t}_{n_s}(x)^H \right) \boldsymbol{\varpi_{n_s}} \tau_{n_s} \nonumber \\
\quad + \sum\limits_{\substack{n_s' \in \mathcal{N} \\ n_s' \neq n_s}} \left( \bm{k}_{n_s}^H \boldsymbol{\Omega}_s \bm{H}^H + \bm{t}_{n_s}(x)^H \right) \boldsymbol{\varpi_{n_s'}} \tau_{n_s'} 
+ \varepsilon_{n_s},$
where \(\bold{\varpi_{n_s}}\) denotes the transmit beamforming vector, \( \tau_{n_s} \) is the information symbol for user \( n_s \) with normalized power, and \( \varepsilon_{n_s} \sim \mathcal{CN}(0, \sigma_{{n_s}}^2)\) represents the additive white Gaussian noise at user \( n_s \),  
which follows a circularly symmetric complex Gaussian distribution. Based on this, the SINR for user \( n_s \) is defined as $\bm{\Gamma}_{n_s} = \frac{\left| \left( \bm{k}_{n_s}^H \boldsymbol{\Omega}_s \bm{H}^H + \bm{t}_{n_s}(x)^H \right) \boldsymbol{\varpi}_{n_s} \right|^2}
{\sum\limits_{\substack{n_s' \in \mathcal{N}, \\ n_s' \neq n_s}} \left| \left( \bm{k}_{n_s}^H \boldsymbol{\Omega}_s \bm{H}^H + \bm{t}_{n_s}(x)^H \right) \boldsymbol{\varpi}_{n_s'} \right|^2
 + \sigma_{n_s}^2} $

The achievable rate for user \( n_s \in \mathcal{N} \), considering short block length, can be expressed as \cite{Jalali2023} 
 \begin{equation}
 R_{n_s}(\bm{x}, \boldsymbol{\varpi_{n_s}}, \boldsymbol{\Omega}_s) = \mathcal{F}_{n_s}(\bm{x}, \boldsymbol{\varpi_{n_s}}, \boldsymbol{\Omega}_s) - \mathcal{G}_{n_s}(\bm{x}, \boldsymbol{\varpi_{n_s}}, \boldsymbol{\Omega}_s)
 \label{equ11}
\end{equation}
where \(\mathcal{F}_{n_s}(\bm{x}, \boldsymbol{\varpi_{n_s}}, \boldsymbol{\Omega}_s) = \log_2(1 + \Gamma_{n_s})\) and
\(\mathcal{G}_{n_s}(\bm{x}, \boldsymbol{\varpi_{n_s}}, \boldsymbol{\Omega}_s) = Q^{-1}(\epsilon_{n_s}) \sqrt{\frac{1}{m_d} \mathcal{V}_{n_s}}\).

In the above expressions, \( \epsilon_{n_s} \) is the decoding error probability of user \( n_s \)and \( m_d \) is the block length, i.e., the number of channel uses,\( \mathcal{V}_{n_s} \) denotes the channel dispersion, given by $\mathcal{V}_{n_s} = a^2 \left( 1 - \frac{1}{(1 + \Gamma_{n_s})^2} \right), \quad \text{where } a = \log_2(e).$
   
\vspace{-8pt}
\section{Problem Formulation}
In this paper, our objective is to maximize the achievable rate for each user \( n_s \in \mathcal{N} \), considering the short block length. This is achieved by jointly optimizing multiple optimization variables, including the surface shape of the FIM antenna, the transmit beamforming vector, the phase shift matrix of the STAR-BD-RIS, and the SINR for each constraints. The corresponding optimization problem is formulated as follows:
\vspace{-10pt}
\begin{subequations} \label{eq:P1}
\begin{align}
\text{(P1)}\max_{\bm{x}, \bm{\varpi_{n_s}}, \boldsymbol{\Omega}_s} \quad & \sum_{n_s=1}^{N} R_{n_s}(\bm{x}, \bm{\varpi_{n_s}}, \boldsymbol{\Omega}_s) \label{eq:15a} \\
\text{s.t.} \quad 
& \Gamma_{n_s} \geq \Gamma_{n_s, \text{min}}, \quad \forall n_s \in \mathcal{N}, \label{eq:15b} \\
& \sum_{n_s = 1 }^{N} \left|\bm{\varpi_{n_s}}\right|^2 \leq P_{max}, \label{eq:15c} \\
& 0 \leq x_p \leq x_{\text{max}}, \quad \forall p \in \mathcal{P}, \label{eq:15d} \\
& \sum_{s \in \mathcal{S}} \left| \Omega_{s}^{f} \right|^2 = 1, \quad \forall f \in \mathcal{F}. \label{eq:15e}
\end{align}
\end{subequations}
The constraint \ref{eq:15b} ensures that the received SINR at each user meets the minimum QoS threshold.  
constraint \ref{eq:15c} imposes a total transmit power limit at the BS. Constraint \ref{eq:15d} restricts the deformation of each FIM surface element to remain within the physical limits.  
The constraint \ref{eq:15e} ensures the unit power operation in the STAR-BD-RIS by imposing a normalization condition on the phase shift vectors. Problem \ref{eq:P1} is an optimization problem with a non-convex objective function and nonlinear constraints, making it a challenging mixed-integer nonlinear programming formulation. Such problems are generally intractable for conventional optimization-based methods. To address this issue, we adopt a Meta-TD3 based DRL framework, detailed in the following section.
\vspace{-12pt}
\section{Problem Solution}

Considering the non-convexity of the proposed optimization problem under system constraints, conventional convex optimization methods are unsuitable, as the high complexity and the extensive search required to find the optimal solution make them unreliable for real-time optimization. In addition, conventional DRL methods may perform poorly in highly dynamic environments due to training–deployment mismatches in next-generation networks. To overcome these challenges, we adopt a meta-enhanced DRL framework and employ the Meta-TD3 algorithm, which improves adaptability and efficiency in complex and dynamic scenarios. Specifically, TD3 efficiently optimizes the problem variables based on markov decision process transitions, while meta-learning adaptively tunes its parameters to enhance convergence speed and robustness. Meta-TD3 preserves the core TD3 structure and incorporates a learnable meta-critic to guide actor updates, thereby improving both convergence and generalization. Building on this, TD3 enhances the baseline deep deterministic policy gradient (DDPG) algorithm with three major techniques:
\begin{enumerate}
    \item \textbf{Use of two critics:} TD3 utilizes two separate critic networks and calculates the Q-value by considering the minimum output from both critics. This approach is beneficial for several reasons; DDPG tends to overestimate the Q-value. By using two critics and selecting the minimum value, TD3 mitigates this overestimation, which helps prevent suboptimal decisions.
    
    \item \textbf{Adding noise to target actions:} When learning the Q-value of target actions, TD3 introduces a small controlled amount of noise to the suggested target actions. This mechanism results in:
    \begin{itemize}
        \item A smoother and more continuous learning process for the critic networks.
        \item Increased robustness against small fluctuations in the state-action space, making the model more resilient.
    \end{itemize}
    
    \item \textbf{Delayed updates for the actor:} In TD3, the actor is updated less frequently than the critics. The rationale behind this delay is:
    \begin{itemize}
        \item If the critics are not yet well-trained, the actor’s learning could be misdirected.
        \item By updating the actor less frequently, TD3 allows the actor to make more informed decisions based on more stable and accurate evaluations from the critics \cite{Mohammadzadeh2024}.
    \end{itemize}
\end{enumerate}


Hence, state set, action set, and reward can be express as: 
\begin{itemize}
    \item {State Space ($S$):}
The state space includes the user channels, the SINR of each user, and the overall system data rate $R_{n_s}$. It can be represented as
\(
S = \Big\{ \{\boldsymbol{\Gamma}_{n_s}, R_{n_s},\bm{k}_{n_s}(x), \bm{t}_{n_s} \}_{ \forall n_s \in N },\{ \bm{h}_f(x) \}_{ \forall f \in F }\Big\}.
\)

\item {Action Space ($A$):}
The action space is composed of all optimization variables in \ref{eq:P1}. Formally, it is given by
\(
A = \big\{\bm{x}, \{\bm{\varpi}_{n_s}\}_{\forall n_s \in N},\{\boldsymbol{\Omega}_s\}_{\forall s \in S} \big\}.
\)

\item{ Reward Function ($rew$):}
The reward function is based on the objective of problem \ref{eq:P1}. Therefore, it can be expressed as
\vspace{-12pt}
\begin{equation}
rew =
\begin{cases}
\displaystyle \sum_{n_s=1}^{N} R_{n_s},&\text{if \eqref{eq:15b}--\eqref{eq:15e} are satisfied}, \\[8pt]
\displaystyle -\sum_{n_s=1}^{N} R_{n_s}, & \text{otherwise}.
\end{cases}
\end{equation}

\end{itemize}

We consider a standard markov decision process, defined by the tuple $(s(t), a(t), \mathcal{P}, rew(t), \gamma)$, where $s(t)$ and $a(t)$ denote the state and action spaces, $\mathcal{P}$ is the probability of state transition, $rew(t)$ is the reward function, and $\gamma \in [0,1)$ is the discount factor. The agent’s objective is to learn a deterministic policy \(\pi_\phi: \mathcal{S} \to \mathcal{A}\) that maximizes the expected cumulative discounted reward $J(\phi) = \mathbb{E}{s_0 \sim \rho_0} \left[ \sum_{t=0}^\infty \gamma(t) rew(s(t), \pi_\phi(s(t))) \right]$

The Q-function or action-value function, evaluates the expected return of taking action a in state $s(t)$ and following policy $\pi$ thereafter. It is formally defined as $Q^\pi(s, a) = \mathbb{E}\pi \left[ \sum_{t=0}^{\infty} \gamma(t) rew(s(t), a(t)) \mid s(0) = s, a(0) = a \right].$
TD3 uses two independent Q-networks, $Q_{\psi_1}$ and $Q_{\psi_2}$ and the smaller Q-value is chosen during target estimation to reduce overoptimistic value estimates, to estimate this function and mitigate the overestimation commonly seen in single-critic approaches.

For each sampled transition $(s(t), a(t), rew(t), s(t+1))$ from the replay buffer, the next action is computed using the target actor and perturbed by clipped noise $\tilde{a}(t+1) = \pi_{\phi’}(s(t+1)) + \epsilon, \quad \epsilon \sim \text{clip}(\mathcal{N}(0, \sigma), -c, c).$
The critic target is computed as $y = rew(t)+ \gamma \cdot \min_{j=1,2} Q_{\psi_j’}(s(t+1), \tilde{a}(t+1)).$
The loss for each critic network is the mean squared error between the predicted and target Q-values $\mathcal{L}_{\text{critic}{(\psi_j})} = \frac{1}{\lvert B \lvert}\sum_{i=1}^ B\left(Q_{\psi_j}(s_i(t), a_i(t)) - y_i \right)^2, \quad j = 1, 2.$
The actor is updated to maximize the expected Q-value from
The first critic $\mathcal{L}_{\text{actor}(\varsigma)} = \frac{1}{\lvert B \lvert}\sum_{i=1}^ B Q_{\psi_1}(s_i(t), a_i(t)) .$
Target networks for both the critics and the actor are softly updated using exponential moving averages:
\begin{equation}
\theta_j’ \leftarrow \tau_1 \theta_j + (1 - \tau_1)\theta_j’, \quad \varsigma’ \leftarrow \tau_2 \varsigma + (1 - \tau_2)\varsigma’, \quad j = 1, 2.
\end{equation}
Here, $\tau_1$ and $\tau_2$ are decay factors that control the updating of the actor and the target of the critic.
To further enhance policy optimization, we incorporate a meta-critic network that learns to guide the actor through an auxiliary gradient.
The critic’s parameters are adjusted by minimizing. The meta-learning procedure can be expressed as a hierarchical, two-stage optimization problem given by:
\begin{equation}
\varkappa = \underset{\varkappa}{\operatorname{arg\,min}} \;
\mathcal{L}_{\text{meta}}\big(b_{\text{val}}; \phi^\ast \big)
\end{equation}
\[
\text{s.t.} \quad
\phi^\ast = \underset{\phi}{\operatorname{arg\,min}}
\left[ \mathcal{L}_{\text{critic}}\big(b_{\text{trn}}; \phi\big)
+ \mathcal{L}_{\text{mcritic}}^\varkappa \big(b_{\text{trn}}; \phi\big) \right]
\]
where $b_{\text{trn}}$ and $b_{\text{val}}$ represent distinct transition batches from the replay buffer for training and validation, $\varkappa$ represents the learned meta-parameters aimed at reducing the meta-loss, defined as \(\mathcal{L}_{\text{meta}} = \tanh \big( \mathcal{L}(b_{\text{val}}; \phi_{\text{new}}) - \mathcal{L}(b_{\text{val}}; \phi_{\text{old}}) \big)\). The update of the actor parameters is carried out in two stages. first, using only the critic loss to produce an intermediate set of parameters
\(\phi_{\text{old}} = \phi - \eta \, \nabla_\phi \mathcal{L}_{\text{critic}}\),
and then applying the meta-critic loss to refine the actor
\(\phi_{\text{new}} = \phi_{\text{old}} - \eta \, \nabla_\phi \mathcal{L}_{\text{mcritic}}^\varkappa\).

The meta-critic parameters are then updated as:
\begin{equation}
\phi \leftarrow \phi - \eta \left( \nabla_\phi \mathcal{L}_{\text{critic}} + \nabla_\phi \mathcal{L}_{\text{mcritic}}^\varkappa  \right).
\end{equation}
\begin{equation}
    \varkappa \leftarrow \varkappa - \eta \, \nabla_\varkappa \mathcal{L}_{\text{meta}}.
\end{equation}

This bi-level optimization ensures that the meta-critic learns to provide feedback that accelerates and stabilizes the actor's policy learning process \cite{farhadi2025joint}.
\vspace{-12pt}
\section{Simulation Results}
\begin{figure*}[!t]
	\centering
 \newcommand{\imgheight}{0.75\linewidth}
 \newcommand{\imgwidth}{\linewidth}      

	\begin{minipage}[t]{0.22\textwidth}
		\centering
		\includegraphics[trim=8cm 0.9cm 3.5cm 3.5cm, clip, width=\imgwidth, height=\imgheight]{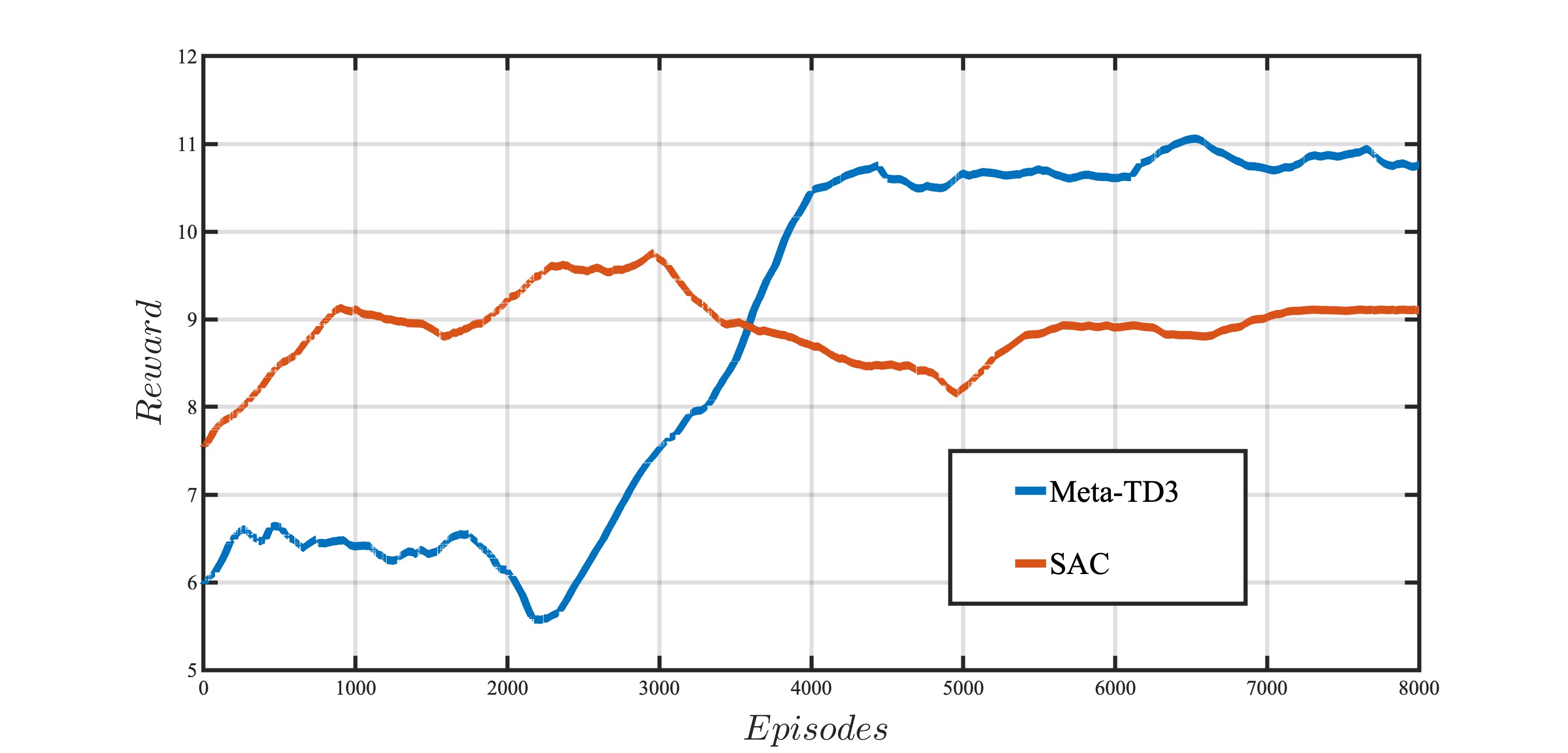}
		\caption{\small Convergence of the proposed system (Meta-TD3 reward curves).}
		\label{Fig_a}
	\end{minipage}%
	\hfill
	\begin{minipage}[t]{0.22\textwidth}
		\centering
		\includegraphics[trim=8cm 0cm 3.5cm 3.5cm, clip, width=\imgwidth, height=\imgheight]{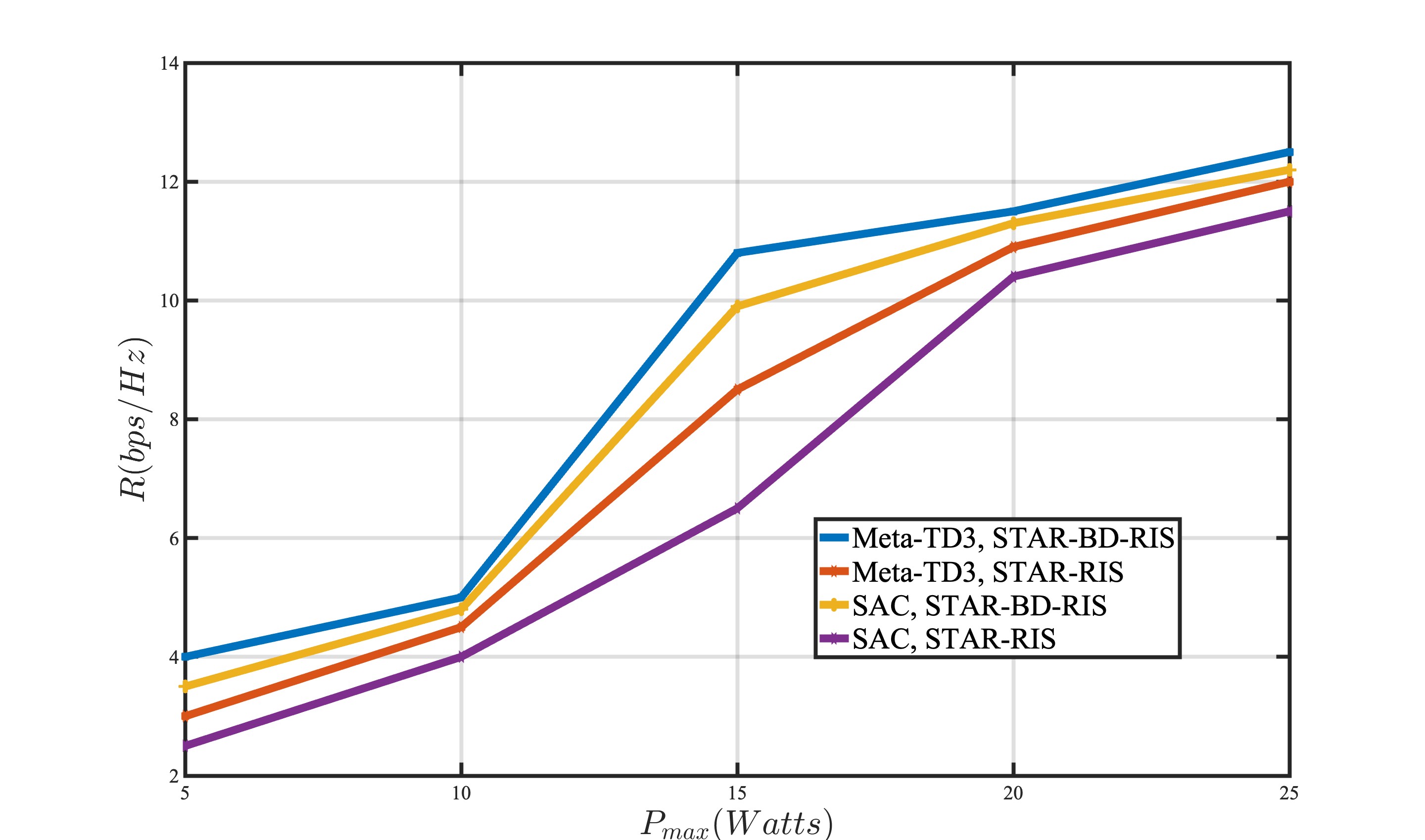}
		\caption{\small Sum rate performance versus transmit power under different algorithms and RIS settings.}
		\label{Fig_b}
	\end{minipage}%
	\hfill
	\begin{minipage}[t]{0.22\textwidth}
		\centering
		\includegraphics[trim=8cm 0cm 3.5cm 3.5cm, clip, width=\imgwidth, height=\imgheight]{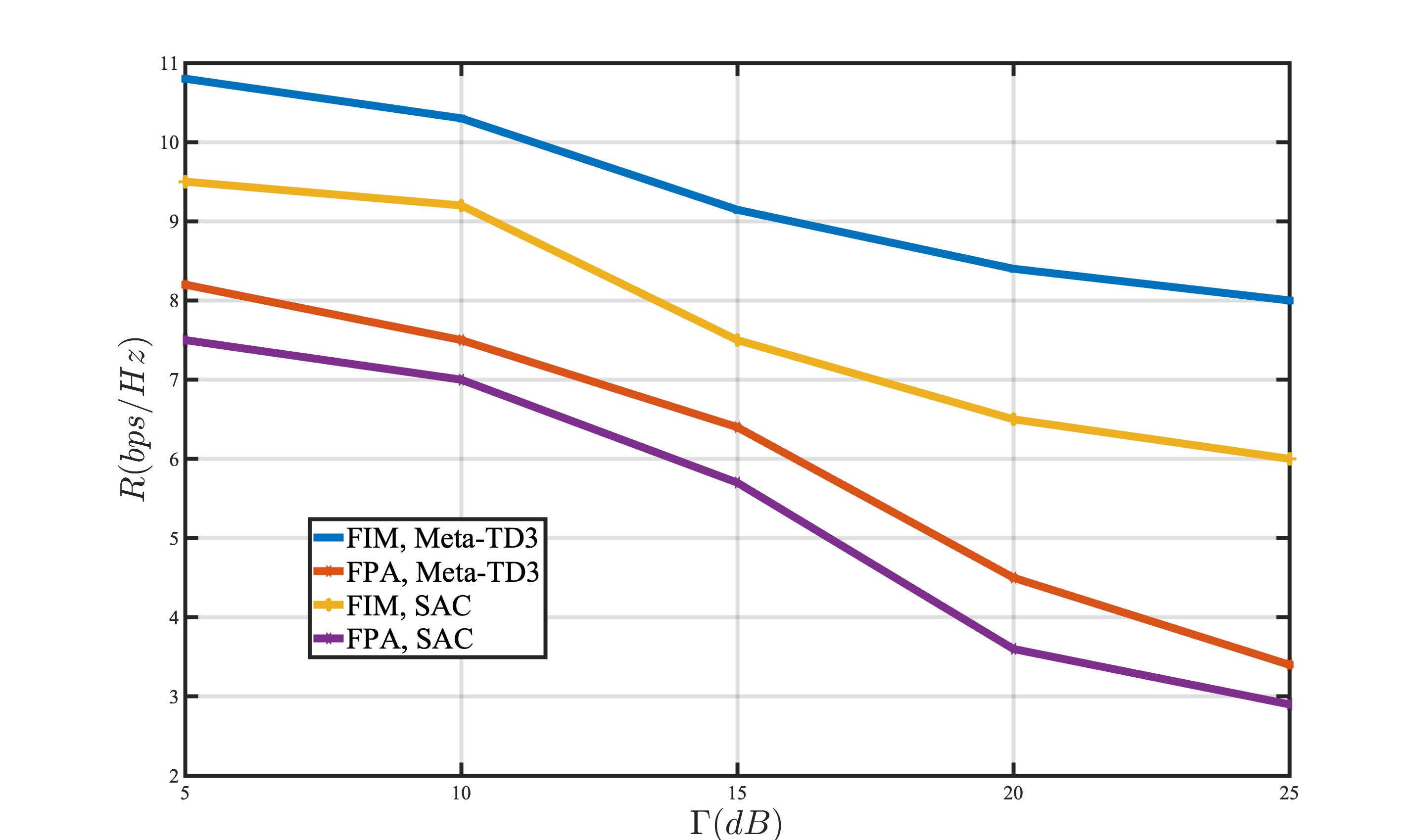}
		\caption{\small Sum rate performance versus SINR under finite block length for FIM and FPA with Meta-TD3 and SAC.}
		\label{Fig_c}
	\end{minipage}%
	\hfill
	\begin{minipage}[t]{0.22\textwidth}
		\centering
		\includegraphics[trim=5cm 0cm 0cm 0cm, clip, width=\imgwidth, height=\imgheight]{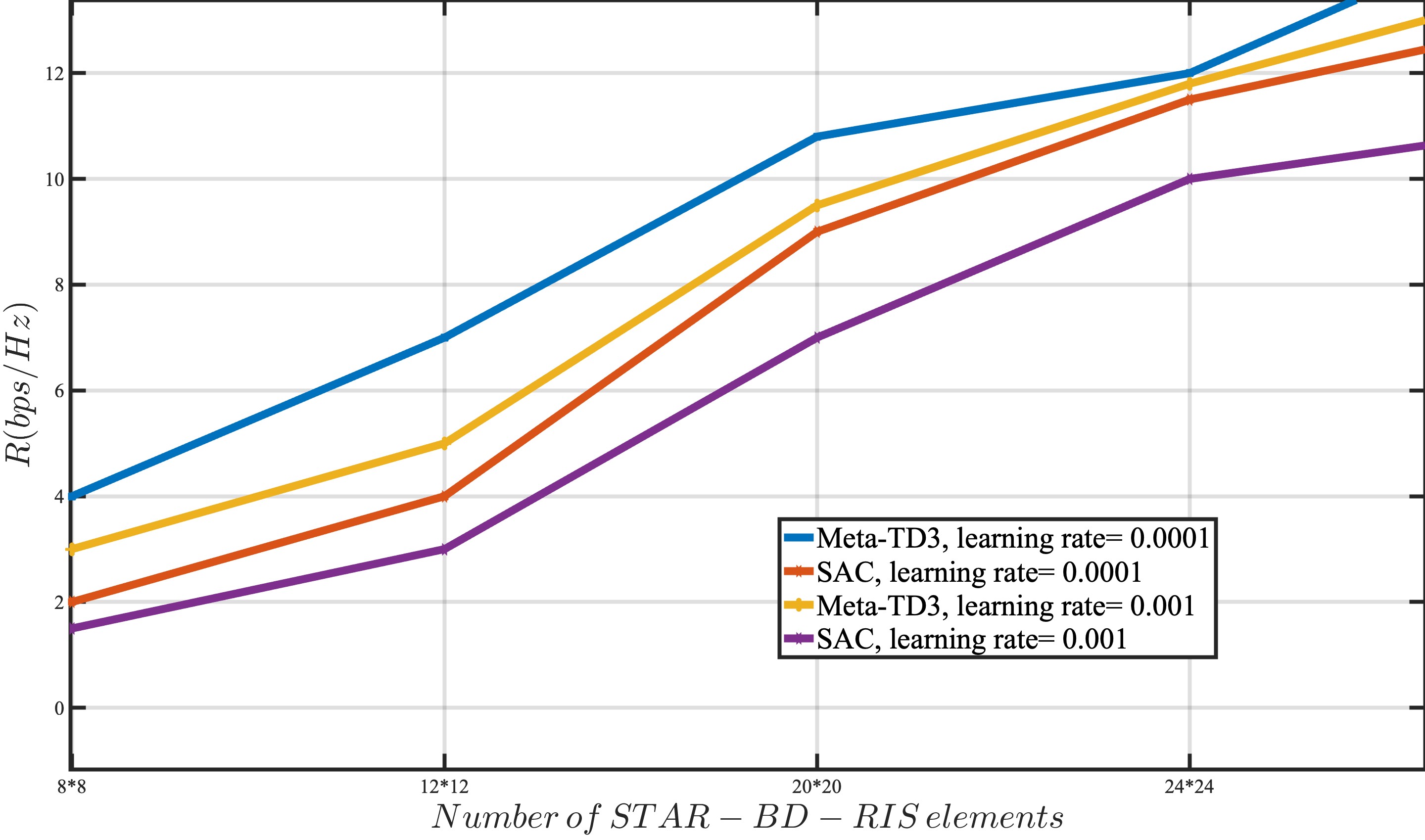}
		\caption{\small Sum rate versus number of STAR-BD-RIS elements for different learning rates (Meta-TD3 and SAC).}
		\label{Fig_d}
	\end{minipage}
	
	\vspace{-20pt} 
\end{figure*}

In this section, we evaluate the performance of the proposed Meta-TD3 algorithm to optimize the joint FIM and STAR-BD-RIS system. All simulation parameters are summarized in Table~\ref{tab1}. The convergence behavior of Meta-TD3 is illustrated in Fig.~\ref{Fig_a}, where it achieves a higher reward compared to the classical soft actor critic (SAC) algorithm. For performance evaluation, SAC is considered as the baseline. The results clearly show that Meta-TD3 significantly outperforms SAC. Although Meta-TD3 requires more episodes to converge compared to SAC, it eventually achieves a higher cumulative reward. This improvement comes from the meta-critic network in our proposed Meta-TD3. In contrast, while SAC converges faster, it suffers from poor generalization and limited adaptability.

\begin{table}
\begin{center}
\caption{\small Simulation Parameters}
\label{tab1}
\begin{tabular}{| c | c | c | c |}
\hline
Parameter & Value & Parameter & Value\\
\hline
$P$ & 16 & Batch Size & 64\\
\hline
$F$ & $20 \times 20$ & Hidden layer1 dim & 500\\
\hline
$N$ & 6 & Hidden layer2 dim & 400\\
\hline 
$D$ & 16 & Hidden layer3 dim & 300\\
\hline
$\gamma$ & 0.99 & $\sigma_{n_s}^2$ & $-22.2~\text{dBm/Hz}$\\
\hline
Learning Rate & 0.0001 & Noise Clip & 0.5 \\
\hline
\end{tabular}
\end{center}
\vspace{-20pt}
\end{table}
\setlength{\textfloatsep}{15pt}
As Fig.~\ref{Fig_b} shows that the joint deployment of FIM, STAR-BD-RIS, and Meta-TD3 achieves superior sum rate performance compared to other benchmarks under the same transmit power limitation. This improvement comes from two key factors: (1) the advanced beam control flexibility of the STAR-BD-RIS architecture, which exceeds that of conventional STAR-RIS designs, and (2) the enhanced adaptability and generalization capability of Meta-TD3 in dynamic wireless environments enabled by meta-learning.
Fig.~\ref{Fig_c} shows the impact of varying SINR thresholds on the performance of the sum rate. 
In addition, it is evident that the FIM antenna equipped configurations consistently outperform those that employ the traditional fixed position antenna (FPA) structure, highlighting the effectiveness of the reconfigurable surface of FIM to improve signal quality. Moreover, the proposed Meta-TD3-based solution achieves superior performance compared to the SAC-based benchmark under both antenna configurations.
As shown in Fig.~\ref{Fig_d}, the impact of the number of STAR-BD-RIS elements on the achievable sum rate is examined under different reinforcement learning settings. It is evident that increasing the number of STAR-BD-RIS elements results in a higher sum rate. This is because a larger number of elements provides the system with greater degrees of freedom to shape the beam more effectively. Consequently, it enables more accurate beamforming, enhances the received signal quality, and mitigates interference, which collectively contribute to an improved sum rate. From the comparison of different investigated learning rates, we can infer that choosing a suitable learning rate can improve the results.

\vspace{-15.5pt}

\section{Conclusion}
In this paper, we investigated a novel system model in which a FIM is equipped at the BS and a STAR-BD-RIS is integrated along the transmission path to enhance coverage and overcome challenging channel conditions. A joint optimization problem was formulated by considering short block length and constraints on the minimum SINR for each user, the maximum available power at the BS, and the parameters of the FIM and STAR-BD-RIS. To efficiently solve this non-convex problem, a Meta-TD3 algorithm was proposed. Simulation results demonstrated that the proposed approach significantly outperforms benchmarks in terms of sum rate performance.

\vspace{-15pt}
\bibliographystyle{IEEEtran}
\bibliography{Bibliography1}

\end{document}